\documentclass[sigconf]{acmart}

\setcopyright{none}

\usepackage{graphicx}
\usepackage[utf8]{inputenc}
\usepackage{algorithmic}
\usepackage{color,soul}
\usepackage{textcomp}
\usepackage{xcolor}
\usepackage{wrapfig}
\usepackage{url}
\usepackage{array}
\usepackage{multirow}
\usepackage{pifont}
 \usepackage{enumitem}

\setlist[itemize,1]{leftmargin=*}
\newcommand{\perspective}{\textsf{PerspectiveCoach}}
\newcommand{\qquote}[1]{{\small\textit{\textcolor{darkgray}{#1}}}}


\begin{document}

\title{\perspective{}: Exploring LLMs for Developer Reflection}
\acmConference[International Conference on Software Engineering SEIS]{International Conference on Software Engineering Software in Society}{April 12-18}{Rio de Janeiro}
\acmYear{2026}
\copyrightyear{2026}

\author{Lauren Olson}
\affiliation{%
  \institution{Vrije Universiteit Amsterdam}
  \country{The Netherlands}
}
\email{l.a.olson@vu.nl}
\author{Emitzá Guzmán}
\affiliation{%
  \institution{Vrije Universiteit Amsterdam}
    \country{The Netherlands}
}
\email{e.guzmanortega@vu.nl}

\author{Florian Kunneman}
\affiliation{%
  \institution{Utrecht University}
    \country{The Netherlands}
}
\email{f.a.kunneman@uu.nl}

\renewcommand{\shortauthors}{Olson, Guzmán, and Kunneman}

\begin{abstract}
Despite growing awareness of ethical challenges in software development, practitioners still lack structured tools that help them critically engage with the lived experiences of marginalized users. This paper presents \perspective{}, a large language model (LLM)-powered conversational tool designed to guide developers through structured perspective-taking exercises and deepen critical reflection on how software design decisions affect marginalized communities. Through a controlled study with 18 front-end developers (balanced by sex), who interacted with the tool using a real case of online gender-based harassment, we examine how \perspective{} supports ethical reasoning and engagement with user perspectives. Qualitative analysis revealed increased self-awareness, broadened perspectives, and more nuanced ethical articulation, while a complementary human–human study contextualized these findings. Text similarity analyses demonstrated that participants in the human-\perspective{} study improved the fidelity of their restatements over multiple attempts, capturing both surface-level and semantic aspects of user concerns. However, human-\perspective{}'s restatements had a lower baseline than the human-human conversations, highlighting contextual differences in impersonal and interpersonal perspective-taking. Across the study, participants rated the tool highly for usability and relevance. This work contributes an exploratory design for LLM-powered end-user perspective-taking that supports critical, ethical self-reflection and offers empirical insights (i.e., enhancing adaptivity, centering plurality) into how such tools can help practitioners build more inclusive and socially responsive technologies.
\end{abstract}

\begin{teaserfigure}
    \centering
    \includegraphics[width=0.43\linewidth]{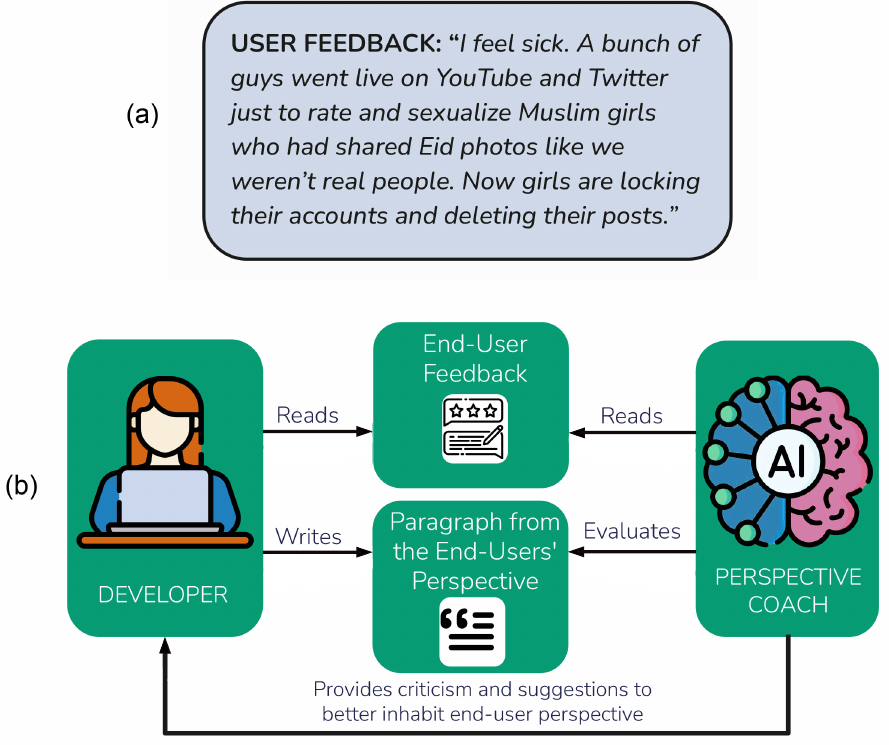}
    \includegraphics[width=0.33\linewidth]{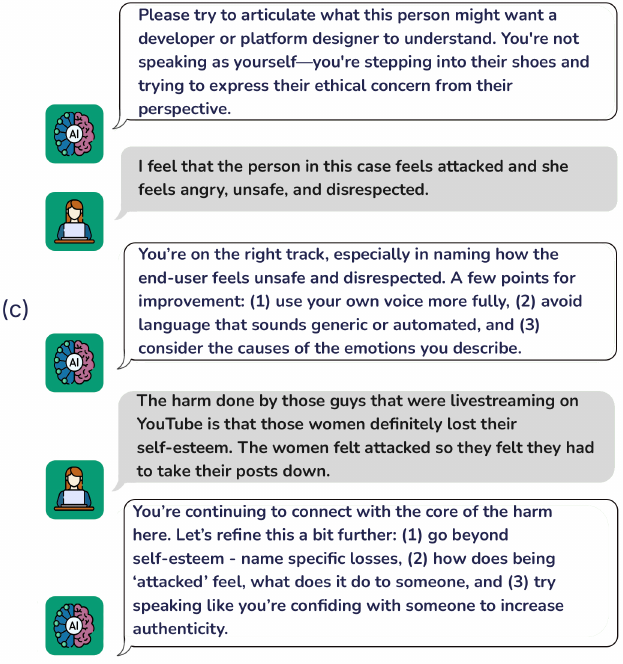}
    \caption{(a) User feedback describing the harassment and objectification of Muslim women during Eid. (b) A schematic overview of the \perspective{}'s reflection pipeline. (c) An example conversation in \perspective{} (summarized from transcript) in which a developer responds to (a).}
    \label{fig:design+convo}
\end{teaserfigure}

\maketitle
\section{Introduction} 
Software users are diverse, globally distributed, and experience a vast range of \textit{ethical concerns}~\cite{tjikhoeri2024best} when interacting with software applications such as misinformation, discrimination, and censorship~\cite{elias2025,kieft2025}. Yet, the software shaping these experiences is predominantly designed by a narrow segment of the global population. Most developers responsible for designing and curating these platforms are white, middle- to upper-class, cisgender, heterosexual, English-speaking men from the United States~\cite{costanza2020design}. Prior work shows that developers’ political orientations can significantly influence their design decisions~\cite{costanza2020design}, meaning the social location of those building technology profoundly shapes what is built, for whom, and with what consequences.

Although some existing research acknowledges the power asymmetries between technology workers and the communities they serve~\cite{ethicstoolkit2024, microsoftResponsibleInnovation2024, googleModelCards2019, c2paHarmsModeling2023}, practical methods to address these imbalances remain limited~\cite{wong2023seeing}. Efforts to “solve” the empathy gap through immersive media, affective interfaces, or bias training have often failed to shift underlying power relations, centering privileged actors’ feelings rather than redistributing epistemic authority~\cite{nakamura2020feeling,messeri2024land}. As Hollanek et al.~\cite{hollanek2024ethico} argue, tools must help developers {\qquote{“reflect on who [they] are instead of pretending they are [the user] and know what [the user] need[s].”}} Our goal is therefore not simply to make developers more empathetic. Instead, we aim to support deeper, more reflective engagement with user feedback that can inform design reasoning across requirements, iteration, and post-deployment contexts, to promote respectful engagement with marginalized communities, and to avoid tokenistic or extractive practices. Such reflection helps teams secure the time, budget, and institutional support needed to authentically serve underserved users and to appreciate the contributions of diverse team members, particularly when those contributions draw on lived experience. 

One promising stance for reorienting design practice in this way is \textit{epistemic humility}~\cite{hollanek2024ethico,smith2021decolonizing}: an awareness of the partial, constructed, and perspectival nature of one’s knowledge~\cite{matthews2006epistemic}. Individuals who exhibit higher epistemic humility are more open to differing viewpoints~\cite{porter2018intellectual}, and empirical research shows that perspective-taking exercises can actively enhance it~\cite{kotsogiannis2024effects}. Building on these insights, we take an exploratory step by introducing \textbf{\perspective}, a Custom GPT designed to guide software developers through structured perspective-taking exercises to deepen developers' critical reflection on how software design decisions affect marginalized users. Our goal is not to present a finalized solution but to explore what an \textit{LLM-based tool} might look like as a first step toward addressing entrenched power imbalances between developers and the communities they impact. 

A critical part of this exploration involves analyzing how conversational dynamics shape perspective-taking. To investigate this dimension, we complement our human–AI evaluation with a human–human study that examines how power, status, prior working relationships, and social identity shape the depth and balance of perspective-taking. We also compare the \textit{restatement quality} of participants’ written responses in both settings using four complementary text similarity metrics (TF--IDF, chrF++, ROUGE-L, and SBERT), allowing us to explore how faithfully participants captured another’s perspective and how this changed across attempts. Our contribution is exploratory: we provide early empirical evidence of how LLM-based perspective-taking might support ethical reflection in software practice, and identify concrete design challenges and opportunities for future iterations of such tools.

\perspective{} extends prior work on ethics scaffolding~\cite{gray2023scaffolding} and reflective design~\cite{sengers2005reflective} by contributing qualitative insights into how developers experience AI-facilitated reflection, including perceived benefits and pain points, as well as how these experiences compare to human–human interactions. In doing so, our work presents early evidence of how LLM-based perspective-taking might be refined to support engagement with marginalized perspectives in software design practice. 

To guide our investigation, we focus on three research questions:

\textit{\textbf{(RQ1)} To what extent does \perspective{} support deeper reflection on design decisions and alternative perspectives, particularly those from marginalized users?} 

\textit{\textbf{(RQ2)} How do developers perceive the usability and relevance of \perspective{} in their design practice?}

\textit{\textbf{(RQ3)} How does developers’ engagement with \perspective{} compare to human–human conversations in terms of conversational dynamics and perspective-taking?} 

To answer these questions, we conducted a controlled study with 18 professional software developers, balanced by sex and screened for relevant front-end design experience. Participants engaged in a multi-turn dialogue with \perspective{}, and rated its usability. Each participant responded to a real example of user feedback describing the online public harassment and objectification of Muslim women on Eid in 2021~\cite{olson2023along,newslaundry2021liberaldoge,bbc2021sullideals,theprint2021consent} and received personalized feedback from the tool. Usability scores for the tool were high across the board, suggesting developers found \perspective{} both engaging and relevant to their work. We complemented this with qualitative open coding of participants’ written reflections, which revealed that many experienced increased self-awareness, broadened perspective, and improved ability to articulate ethical reasoning. To contextualize these findings, we also conducted a human–human perspective-taking study that examined how power, identity, and prior working relationships shape conversational dynamics. Analysis of turn-taking patterns and perspective restatement quality showed that power asymmetries and lack of established relationships were associated with conversational imbalance and reduced perspective-taking, highlighting key areas for future tool design. We include a replication package to encourage that further work.\footnote{\url{doi.org/10.6084/m9.figshare.30231535}}
\section{Related Work}

\textbf{Power Asymmetries and the Limits of Empathy Technologies}

A growing body of work has documented how power asymmetries shape the design and governance of software systems, often marginalizing the needs and voices of those most affected by them~\cite{ethicstoolkit2024, microsoftResponsibleInnovation2024, googleModelCards2019, c2paHarmsModeling2023, wong2023seeing}. Scholars have highlighted that most software is built by a demographically narrow group—predominantly white, cisgender, English-speaking men from the Global North~\cite{costanza2020design}—which contributes to epistemic monocultures that fail to capture diverse user experiences. While some industry and academic initiatives aim to unveil harms and address inequities, practical methods to redistribute epistemic authority remain limited. Recent SE work documents how subtle micro-inequities shape who feels able to speak and be heard in software contexts, including immigration-linked dynamics in industry settings and gendered disparities in everyday work practices~\cite{markulj2024micro,guzman2024mind}.

Attempts to resolve these imbalances have often turned to empathy-building technologies. From early utopian narratives about the internet’s democratizing potential and identity experimentation in online role-playing games~\cite{turkle2011life}, to more recent efforts to automate compassion through virtual reality experiences~\cite{nakamura2020feeling} and therapy-oriented chatbots, new technologies have repeatedly been framed as tools to foster understanding and social change. However, as Nakamura and Messeri argue, such interventions risk centering privileged actors’ feelings rather than empowering marginalized groups ~\cite{nakamura2020feeling, messeri2024land}. Our work extends these critiques by focusing not on cultivating emotional empathy per se, but on designing tools that help developers engage in non-tokenistic, respectful forms of collaboration, ones that foreground marginalized perspectives and enable those most affected by technology to lead conversations about its impacts.

\textbf{Reflection, Epistemic Humility, and Values Work}
Within HCI, several streams of research have explored approaches to integrate ethical reflection and values into design. Reflective design~\cite{sengers2005reflective} and ethics scaffolding~\cite{gray2023scaffolding} propose methods to challenge assumptions and make value tensions explicit during development. In RE, “Value Stories” were proposed to surface and trace values into requirements artifacts, but primarily as workshop scaffolds rather than reusable, developer-facing tooling~\cite{detweiler2014value}. Scholars have argued that cultivating \textit{epistemic humility}—an awareness of the partiality and situatedness of one’s own knowledge—can help developers engage more ethically with diverse stakeholders~\cite{hollanek2024ethico, smith2021decolonizing, matthews2006epistemic}. Such humility encourages designers to question their assumptions, recognize knowledge gaps, and seek out perspectives beyond their own. Empirical work suggests that perspective-taking exercises can enhance epistemic humility~\cite{kotsogiannis2024effects}, though practical tools to support this process in software design remain nonexistent. A recent SLR in SE catalogs 85 primary studies on ethical values and stakeholders but highlights the fragmentation of methods and the lack of developer-facing operational guidance~\cite{alidoosti2025exploring}.

Our work builds on this foundation by exploring how a conversational agent might scaffold structured perspective-taking as part of ethical reflection. Rather than positioning empathy as an endpoint, \perspective{} aims to foster deeper critical thinking about marginalized experiences and to motivate more proactive forms of engagement, such as co-design and participatory approaches~\cite{spinuzzi2005methodology}, that avoid tokenism and exploitation.

\textbf{Design Practices, Ethnography, and Values Mediation}

Research on values work in UX and software practice has shown that integrating ethical considerations into design is rarely straightforward. Gray and colleagues~\cite{gray2016s, gray2019meditation} show that practitioners often conceptualize ethics as a ``mindset'' rather than a method, mediating between competing pressures in ways that are highly context-dependent. However, these studies also reveal that designers’ interaction with end-users is frequently indirect or mediated, limiting the depth of perspective-taking that occurs in practice. Chivukula et al.~\cite{chivukula2020dimensions} similarly identify organizational, temporal, and relational factors that shape ethical awareness, highlighting how power, time constraints, and institutional logics can constrain values work.

Bridging these gaps has been a persistent challenge. Khovanskaya et al.~\cite{khovanskaya2017reworking} argue that ethnographic insights about values often get lost as they move through organizational pipelines, calling for new strategies to translate contextual knowledge into design action. SE studies increasingly conceptualize empathy as relevant yet underspecified, identifying enablers/barriers in developer–user communication and calling for concrete practices that support empathic work in situ~\cite{gunatilake2024enablers,gunatilake2025role,cerqueira2025exploring}. Complementing empirical work, recent conceptual syntheses map empathy constructs from psychology to SE and argue that SE lacks operationalized, developer-usable mechanisms that connect these constructs to everyday practices~\cite{gunatilake2023empathy}. Our work responds to this call by exploring how LLM-based conversational tools might scaffold values-oriented reflection earlier and more continuously in the software design process, preserving critical user perspectives that might otherwise be flattened or lost.
\section{Tool Design}
\begin{figure}

    \includegraphics[width=1.07\linewidth]{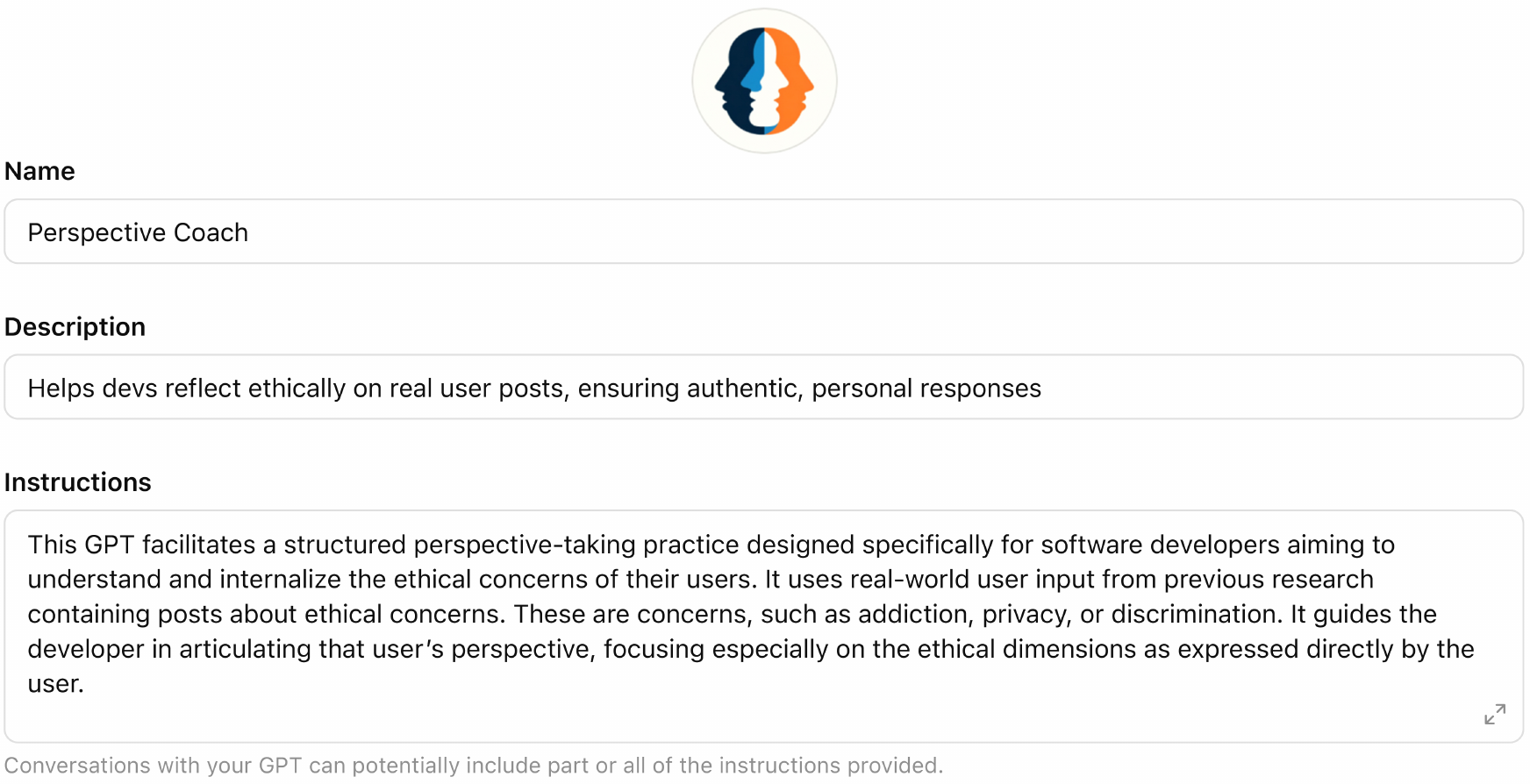}
 \caption{\perspective{}’s configuration interface, where you can define the GPT’s name, description, and behavior. Instructions visible here represent only a subset of the prompt.}
 \Description{\perspective{}’s configuration interface, where you can define the GPT’s name, description, and behavior. Instructions visible here represent only a subset of the prompt.}
    \label{fig:interface}
 \end{figure}

Figure~\ref{fig:design+convo} shows the overall design of \perspective{} and an example of its interactions. The tool is designed to help developers practice perspective-taking by engaging with real user concerns in a supportive, feedback-driven conversation. To ensure usability within developer workflows and effectively facilitate perspective-taking, we deployed the \perspective{} through OpenAI’s Custom GPT platform in May 2025~\cite{openai2023introducinggpts}. This choice offers a well-known web interface that integrates easily into existing development practices. Further, recent studies have shown that GPTs can outperform humans in empathic communication~\cite{kuhbacher2024chatbot,lee2024large}. These models demonstrate strong cognitive empathy and emotion recognition abilities~\cite{schaaff2023exploring}, and can be fine-tuned to deliver emotionally responsive interactions through prompting alone, even in zero-shot settings~\cite{10388177}. Participants accessed the tool through a direct link during the study, and all interactions took place within the default GPT interface. The core functionality of \perspective{} is entirely prompt-driven: we iteratively prompted updates to its system instructions until the key features were reliably expressed (see Fig.~\ref{fig:interface}), consistent with findings that LLM-driven prompt optimization outperforms manual tuning ~\cite{zhou2022large,pryzant2023automatic}. The tool adopts a tone that is thoughtful, constructively critical, and consistently supportive of the user’s learning. Foundational work on formative assessment shows that constructive feedback is critical for learners to deepen understanding~\cite{nicol2006formative}, while studies of psychological safety demonstrate that a consistently supportive environment increases individuals’ willingness to reflect, experiment, and engage with challenging feedback~\cite{edmondson1999psychological}.
 Key features implemented via prompt logic include:

\begin{itemize}
    \item \textbf{Fidelity enforcement:} \perspective{} discourages copying and pasting of the user’s original post. It detects near-verbatim phrasing and prompts participants to rephrase using their own understanding, helping preserve the authenticity of the perspective-taking process.

    \item \textbf{Bias avoidance:}  \perspective{} is directed to avoid introducing its own moral judgments and flags instances where participants insert ethical interpretations or blame not present in the original post. It emphasizes deep listening and discourages developer projection or GPT-based embellishments.

    \item \textbf{Supportive scaffolding:} In line with foundational work on tutoring~\cite{wood1976role}, when participants struggle to articulate the end-user's ethical concern, \perspective{} offers structured support, such as guided rephrasing suggestions and focused prompts, to help them reflect more deeply. `Scaffolding' gradually decreases as learners demonstrate more confidence.
\end{itemize}

While the structure of the task is consistent, the conversation itself is adaptive. Because \perspective{} is built on GPT-4, it responds flexibly to user input, tailoring its feedback to individual reflection quality and offering custom suggestions for revision. There is no formal summary or fixed closing phrase; instead, \perspective{} provides dynamic feedback on users' perspective-taking ability, allowing them to gauge when the conversation has naturally concluded. \textbf{To support replication and transparency, we include the full system instructions used to configure \perspective{} in our replication package.}
\section{\perspective{} Evaluation}

We conducted a mixed-method evaluation to examine how \perspective{} supports developers in engaging with marginalized users’ ethical concerns through structured perspective-taking exercises. Our study combined a controlled human–AI experiment with 18 professional front-end developers and a complementary human–human comparison to contextualize conversational dynamics and meaning-making processes. Through surveys, qualitative coding, text similarity analyses, and turn-taking metrics, we assessed how participants reflected on design decisions, articulated alternative perspectives, and navigated conversational asymmetries. Together, these findings offer empirical insights into \perspective{}’s strengths, current limitations, and key design opportunities for improving adaptive guidance, personalization, and conversational responsiveness.

\subsection{Human-AI Study Setup} This study investigates how software practitioners engage in structured perspective-taking when responding to end-user ethical concerns, particularly in contexts involving gendered harm. Participants are presented with a five-step task sequence conducted through \perspective{}.

Participants began by reading and agreeing to an university-approved informed consent statement outlining the study's purpose, procedure, and data privacy measures. Following this, participants engaged with \perspective{}. After the interaction, they filled out a structured perceived reflection and usefulness survey (see Table~\ref{table:questions}). The first two items (Q1–Q2) were designed to address \textbf{RQ1} by probing whether \perspective{} supported deeper reflection on design decisions and broadened participants’ consideration of alternative perspectives, particularly those of marginalized users. Our selection of Q3 and Q4 emphasizes the tool’s relevance to developers’ day-to-day work (e.g.,: reflecting on design decisions and articulating values-based reasoning), because many ethical tools and frameworks are perceived as too abstract or disconnected from practice~\cite{wong2023seeing, sirur2018we, selbst2019fairness}. Q5 assesses participants’ willingness to reuse the tool, a standard proxy for perceived usefulness and integration potential in practice~\cite{AJZEN1991179}.

All participant responses are anonymous. Participants are asked to assume the role of a software developer at a major social media platform (e.g., YouTube or Twitter). Each participant is provided with a real user post describing an ethical concern gathered in a previous study~\cite{olson2023along}; in this case, an incident involving the public sexualization and harassment of Muslim women during Eid (see Figure~\ref{fig:design+convo} (a))~\cite{newslaundry2021liberaldoge,bbc2021sullideals,theprint2021consent}. We selected this scenario because it represents a harmful incident that was widely documented and criticized in external reporting~\cite{newslaundry2021liberaldoge,bbc2021sullideals,theprint2021consent}, allowing us to verify that the described events occurred: a level of confirmation not possible for most user posts while remaining obscure enough to limit participants' pre-existing beliefs. Participants are instructed to copy the text of the post into the \perspective{} interface and engage in five rounds of guided reflection, where the \perspective{} prompts them to re-articulate the user’s experience in their own words, focusing on the ethical dimensions as described by the original user. Participants took an average of 45 minutes to complete all tasks and were compensated at the rate of around \$USD13/hour.

\subsubsection{Participants}
We recruited 18 participants from Prolific, including 9 female and 9 male respondents. This 50/50 sampling decision was purposeful: males remain overrepresented in software development roles~\cite{EEOC2024} and are statistically less likely to experience sex-based harassment online~\cite{vogels2021state}. We analyze sex because Prolific includes sex instead of gender in its available demographic data~\cite{prolificDemographicExport}. Prolific collects sex based on legal documents, which may not align with participants’ self-identified gender due to legal, social, and transitional complexities. While our analysis reflects socially shaped attributes more than biological ones, we acknowledge the conflation of sex and gender and the interpretive limitations this entails.

To ensure relevant technical background, participants were required to have recent experience with front-end software development activities, including debugging, functional testing, unit testing, responsive design, UI design, A/B testing, or UX work. We verified this experience using a method adapted from Schmidt et al.~\cite{schmidt2023accountability}, in which participants described one of their last three front-end development projects and their specific role in it. Two authors independently reviewed these descriptions to confirm qualification. All participants were also required to be from the United Kingdom (n = 9), the United States (n = 5), Australia (n = 2), Canada (n = 2), or New Zealand (n = 0), due to their Anglocentric, Global North perspective.

\subsubsection{Survey Responses}
Post-interaction survey data were analyzed descriptively to assess the perceived usefulness and usability of \perspective{}. Participants rated five statements on a five-point Likert scale (1 = strongly disagree, 5 = strongly agree) (see Table~\ref{table:questions}). We calculated the mean, standard deviation, minimum, and maximum values for each item to characterize central tendencies and variability in participants’ responses. 
\begin{table*}[h]
\small
\centering
\begin{tabular}{l}
\hline
 \textbf{Question} \\
\hline
 \textbf{Q1:} The \perspective{} helped me reflect more deeply on my own design or development decisions. \\
\textbf{Q2:} The tool helped me better consider alternative perspectives (e.g., user needs, ethical concerns). \\ 
 \textbf{Q3:} I found the feedback and questions from the \perspective{} relevant to my work. \\
 \textbf{Q4:} The tool made it easier to articulate and defend values-based decisions in a technical context. \\
 \textbf{Q5:} I would use a tool like this again when making complex or ethically sensitive decisions. \\ \midrule
 \textbf{Q6:} Please describe (1) what you found helpful about the chatbot if anything and (2) what you found unhelpful about the chatbot if anything. \\
 \textbf{Q7:} What improvements would you suggest for future versions of the tool? \\
\bottomrule
\end{tabular}
\caption{Outline of the survey questions used. Q1-Q5 are Likert scale (1-5) questions and Q6-Q7 are open text questions.}
\label{table:questions}
\end{table*}
\subsubsection{Qualitative Coding of Participant Feedback}
To deepen our understanding of participants’ experiences with \perspective{}, we conducted an inductive qualitative analysis of their post-interaction responses to two open-text questions. Two authors independently open-coded all responses, generating 38 and 64 initial codes from 60 and 125 quotations, respectively. The coding sets were then merged into 42 codes based on shared meanings and overlapping interpretations by the two coders. In a final collaborative session, the two authors organized these codes into seven overarching themes~\cite{saldana2021coding}. 
\subsection{Human-AI Study Results \textit{(RQ1 \& RQ2)} }
\subsubsection{\textbf{\textit{(RQ1)}} Developer Reflection on Decisions \& Perspectives}
Here we address whether \perspective{} meaningfully deepens developers’ reflection on design decisions and alternative perspectives. We examine both perceived impact and concrete reflective behaviors by combining (a) survey questions Q1 and Q2 (see Table~\ref{table:questions}) and (b) open-ended comments analyzed via open coding to capture how reflection actually unfolded for practitioners. This mixed approach lets us gauge not only if participants felt more reflective, but how the tool guided that reflection so we can translate insights into actionable design changes.

\textbf{Survey Response Results.} The highest-rated item (Q2, $M = 4.74$, $SD = 0.45$, $\text{range} = 4$–$5$) suggests that participants felt the tool effectively supported them in considering alternative perspectives, including user needs and ethical concerns. Scores were similarly high for the tool’s ability to deepen reflection on design decisions (Q1, $M = 4.68$, $SD = 0.95$, $\text{range} = 1$–$5$) and to support articulation of values-based reasoning in technical contexts (Q4, $M = 4.68$, $SD = 0.58$, $\text{range} = 3$–$5$). 

\textbf{Open Coding Results.}  

\textit{\textbf{Coach as Teacher:}}
Participants described the chatbot as playing an educative role rather than merely facilitating dialogue. One referred to it as \qquote{“ethically educative,”} with another noting that the iterative process provided \qquote{“clear, concrete feedback and examples [that were] effective in refining my ability to communicate the emotional depth and personal meaning of the user's experience.”} Respondents emphasized that it helped them \qquote{“organize complex emotions into actionable insights” and “break down the problem into actionable steps,”} transforming reflection into more tangible outcomes. Some valued its \qquote{“specific suggestions for areas of improvement”} and described \qquote{“every single thing [as] helpful, especially the precise feedbacks.”}

\textit{\textbf{Connecting with Emotions:}}
Participants highlighted how the tool supported emotional articulation and deeper engagement with difficult feelings. One noted that it prompted them \qquote{“to think further and articulate more precisely what I thought and felt regarding the scenario — and this is something I often glaze over, especially when I am upset.”} Others described improvements in \qquote{“my ability to communicate the emotional depth and personal meaning of the user's experience,”} suggesting the tool effectively scaffolded deeper emotional connection to the design context.

\textit{\textbf{Conversational Guide:}}
Many participants described the chatbot as a helpful guide that kept them focused and deepened reflection. It \qquote{“kept me focused on the topic,”} \qquote{“encouraged me to think further and articulate more precisely,”} and \qquote{“helped me explore the emotions of someone going through such an experience.”} Several emphasized that their views were \qquote{“taken seriously”} and built upon, that the agent \qquote{“made me think in a different way,”} and that it \qquote{“helped me reflect on myself more deeply.”} Participants valued its ability to \qquote{“build on my emotional perspective”} while prompting them to \qquote{“think further”} and sustain engagement with the ethical issue.

\textit{\textbf{Coach Tone:}}
Participants described the chatbot’s tone as generally\qquote{ “clear, concise, and structured,”} with \qquote{“positive reinforcement”} and an \qquote{“encouraging tone when addressing and quoting my writing.”} One appreciated that “\qquote{the bot pushed me}” to deepen engagement, while others noted the process was \qquote{“effectively structured”} to facilitate perspective-taking. However, one participant asked for \qquote{“clearer initial instructions”} on how to interact with the chatbot, suggesting that small onboarding improvements could strengthen early exchanges. Others felt the tone was at times \qquote{“too polished”} or formal: \qquote{“At times, the chatbot could be too focused on providing polished responses instead of reflecting more casually or naturally… A more conversational tone might have felt more natural and engaging for personal reflection.”}

\textit{\textbf{Good Personalisation:}}
Users valued moments when the chatbot \qquote{“captured and built on my emotional perspective”} and \qquote{“expanded” their views.} Participants reported that their perspective felt \qquote{“heard and taken seriously”} and that the chatbot \qquote{“helped me get into the shoes of the person.”} Such personalization contributed to perceptions of empathy and relevance, deepening the sense of being understood and taken seriously by the tool.
\subsubsection{\textbf{\textit{(RQ2)}} Usability \& Relevance for Developer Workflows}
Here we examine how usable and relevant \perspective{} feels within developers’ day-to-day workflows. We combine three survey questions (Q3, Q4, Q5, see Table~\ref{table:questions}) with open-ended feedback analyzed via open coding to surface concrete opportunities and pain points. The section proceeds in two parts: first, summarize perceived usefulness and relevance; second, distill what developers say they want, from writing support and UI refinements to personalization and flexibility.

\textbf{Survey Response Results.} Participants  expressed a strong likelihood of future use (Q5, $M = 4.68$, $SD = 0.58$, $\text{range} = 3$–$5$), underscoring the tool’s perceived relevance and practical value. Although responses to the relevance of feedback (Q3, $M = 4.58$, $SD = 0.96$, $\text{range} = 1$–$5$) showed slightly more variability—with a small number of participants rating this item much lower than the overall trend—the results overall indicate that developers found \perspective{} both engaging and useful for integrating ethical reflection into their design work.

\textbf{Open Coding Results.}

\textit{\textbf{Writing Coach:}}
Beyond perspective-taking, many participants framed the chatbot as a writing coach. It was described as \qquote{“helpful in refining responses,”} \qquote{“helpful in writing higher quality material,”} and valuable for providing \qquote{“tools… in the aspects of rephrasing.”} Several highlighted how it helped them \qquote{“break habits of impersonal writing,”} such as defaulting to third-person voice or overly abstract language, and supported them in \qquote{“improving formulation”} of their responses.

\textit{\textbf{User Interface:}}
While most feedback centered on conversational aspects, several participants offered UI-focused suggestions. Some proposed a \qquote{“softer UI”} to match the coaching metaphor, \qquote{“something less tech and a bit more zen… soften the edges — bring in beige… it needs something gentle.”} Others suggested adding \qquote{“voice interaction support,”} improving speed (\qquote{“it could be faster”}), and expanding features such as the conversation overview: \qquote{“I like it at the end of the chat when the chatbot automatically puts all my responses together, rather than me scrolling up and see the back and forth.”}

\textit{\textbf{More Personalized:}}
Several participants wanted deeper personalization and adaptivity from the chatbot. They asked for \qquote{“alternate versions of feedback” }and noted that \qquote{“some responses… seemed generic as it would repeat as opposed to create.”} Suggestions included \qquote{“making the tool a bit more intuitive in recognizing when deeper exploration is necessary,”} allowing \qquote{“more flexibility to diverge from the script,”} and adapting \qquote{“the tone to match the user’s preference.”} Others called for \qquote{“more work on the feedback”} to reflect that \qquote{“some perspectives are personal”} and requested \qquote{“more emphasis on practical examples,”} such as \qquote{“real-world tech interventions to ground abstract ethics.”} Participants also noted redundancy (\qquote{“it occasionally repeated ideas”}) and urged a \qquote{“balance [between] emotional precision [and] space for unfiltered responses.”} Similarly, a participant described the chatbot as overly directive. They felt that \qquote{“the rigid step-by-step flow felt restrictive”} and limited their ability to pursue tangential or emergent ideas. This was seen as especially limiting when participants wanted to follow their own train of thought rather than remain tightly guided by the chatbot’s structure.

\subsection{Human-Human Study Setup} 
This study aimed to ground \perspective{}’s evaluation in real interactions by examining \textit{thick perspective-taking}, extending the notion of \textit{thick empathy}~\cite{crockett2025empathy} to perspective-taking that requires interpersonal knowledge or shared experience. Thick empathy requires experiential or interpersonal knowledge and is thus the standard yet neglected form of empathy due to the practical difficulty of studying real relationships and lived experience.

Pairs completed an in-person, one-hour perspective-taking session within a software–focused research group.  Participants were purposefully matched to introduce a power dynamic (e.g., PhD–assistant professor), reflecting the kinds of asymmetries the \perspective{} coach is designed to help developers navigate in real-world ethical conversations. After informed consent and a standardized briefing, each participant independently selected all ethical issues they had personally experienced when using software applications (e.g., accessibility, privacy). Pairs compared answers and intentionally chose one concern experienced by only one partner to create an asymmetry of perspective (see pair demographics in Table~\ref{table:hh_results}). P3 was online, the rest in-person. Both partners then wrote a 1–2 paragraph, first-person account describing a concrete incident, feelings, why it mattered, and the specific platform involved. Participants recorded a transcript using software and saved a copy to preserve an unedited record. No compensation was provided.
Each partner then wrote a perspective-taking paragraph re-articulating the other’s experience "in their own words"; these paragraphs were all typed in online documents. Pairs exchanged paragraphs, provided feedback, and iterated until both reported feeling understood (no fixed number of cycles). A facilitator was present to answer questions and keep procedures/timing consistent but did not coach content beyond the written worksheet steps. Collected artifacts included: (1) the initial first-person narrative, (2) the partner’s perspective-taking paragraph, (3) any revisions after feedback, (4) the transcript, and (5) brief post-study comments. 

\subsubsection{Text Similarity Analysis}
To evaluate how faithfully participants rearticulated original user experiences in their written restatements (\textit{Attempt 1}), we conducted a text similarity analysis between each restatement and its corresponding source narrative. Following established practices in computational linguistics~\cite{salton1975vector,manning2008introduction,lin2004rouge,popovic2015chrf,reimers2019sentence}, we selected four complementary metrics that capture distinct dimensions of similarity:

\begin{itemize}
    \item \textbf{TF--IDF cosine similarity}~\cite{salton1975vector,manning2008introduction} measures lexical overlap weighted by term importance, indicating how much of the original vocabulary and topical content was retained.
    \item \textbf{chrF++}~\cite{popovic2015chrf} captures surface-form similarity at character and word n-gram levels, providing sensitivity to phrasing, morphology, and fluency beyond simple word overlap.
    \item \textbf{ROUGE-L}~\cite{lin2004rouge} computes similarity based on the longest common subsequence, reflecting preservation of structural order and sequencing of ideas.
    \item \textbf{SBERT semantic similarity}~\cite{reimers2019sentence} embeds both the original and restated texts into a shared vector space using a transformer-based model and computes cosine similarity between them. Unlike overlap-based metrics, SBERT captures paraphrases and meaning-level equivalence even when surface forms diverge, making it particularly useful for assessing conceptual fidelity.
\end{itemize}

Each metric ranges from 0 (no similarity) to 1 (perfect similarity). Because absolute thresholds for these measures are context-dependent, we interpreted scores relative to the observed distribution in our dataset. For TF–IDF, scores above 0.45 were considered \textit{high}, 0.35–0.45 \textit{medium}, and below 0.35 \textit{low}. For chrF++, \textit{high} was defined as above 0.32, \textit{medium} as 0.20–0.32, and \textit{low} below 0.20. For ROUGE-L, scores above 0.23 were treated as \textit{high}, 0.16–0.23 as \textit{medium}, and below 0.16 as \textit{low}. For SBERT, \textit{high} similarity was defined as above 0.75, \textit{medium} as 0.45–0.75, and \textit{low} below 0.45. Because overlap-based metrics are sensitive to text length, phrasing variation, and reference choice, even semantically faithful paraphrases of short inputs often receive relatively low absolute scores, making within-dataset ranking a reliable indicator of faithfulness to the original text~\cite{popovic2015chrf,lin2004rouge,grusky2023rogue}. We interpret higher scores as indicative of more faithful restatements and thus stronger perspective-taking, while lower scores suggest important omissions, rewording, or shifts in emphasis. Because linguistic similarity does not capture all aspects of interpretive quality, we also examined which participants requested revisions to their restatements. 

\subsection{Human-Human Study Results \textit{(RQ3)}}
We further analyzed conversational dynamics in the human--human study as an additional proxy for perspective-taking quality. Participants shared a range of ethical concerns, with some overlapping with the harassment scenario from the human-AI study. Restatement analyses showed variation in similarity to the original perspective, with lower similarity often prompting revision requests. We also compared initial restatement quality between human-human and human-AI restatements. Because perspective-taking involves not only rephrasing another person’s experience, but also listening, we examined patterns of \textit{turn-taking} and \textit{verbosity} across four participant pairs.

\subsubsection{Content of Participants’ Concerns}

\begin{table}[h]
\centering
\footnotesize
\caption{Summary of participants’ concerns in the human--human study.}
\label{tab:concerns_summary}
\begin{tabular}{ll}
\toprule
\textbf{Speaker} & \textbf{Concern} \\
\midrule
S1 & Privacy: unnecessary requirement to disclose home address \\
S2 & Identity theft: cloned Instagram profile used to solicit money \\
S3 & Limited service access: Spotify restricted outside Europe \\
S4 & Data breach: bank and personal details stolen in Patreon hack \\
S5 & Cyberbullying: hateful comments targeting identity/beliefs \\
S6 & Privacy breach: hacked Google account and threats to leak media \\
S7 & Discrimination: lack of access to popular apps in home country \\
S8 & Sustainability: visible resource waste in online services \\
\bottomrule
\end{tabular}
\end{table}
The ethical concerns participants selected and described in the human--human study reflected a diverse range of experiences with digital technologies (Table~\ref{tab:concerns_summary}). These included issues related to privacy (S1, S6), security breaches (S2, S4), platform discrimination and access inequities (S7), sustainability (S8), and online harassment and abuse (S5). For example, S1 described being required to disclose their home address to a service that did not require it to function, while S6 reported a hacked Google account and subsequent threats to release sensitive personal photos and videos. S2 recounted a case of identity theft on Instagram, where their profile was cloned and used to solicit money and information from others. S4 described the theft of bank and personal data in a Patreon data breach. Other concerns focused on access and equity, such as S3’s inability to use Spotify beyond 15 days outside of Europe and S7’s experience of platform discrimination due to geographic restrictions on widely used apps like Netflix. S8’s concern centered on sustainability, describing frustration at the visible waste of resources across online services. Finally, S5 reported being targeted with hateful comments online based on aspects of their identity and belief system.

Notably, three of these concerns (S5, S6, and S2) shared substantive thematic overlap with the harassment case used in the human--AI study, all centering on the non-consensual use of images or identity-based harassment. S5 involved targeted harassment directed at the participant based on aspects of their identity and belief system, closely mirroring the gendered abuse described in the Eid example. S6 likewise concerned harassment linked to sensitive, private images, echoing the original scenario’s dynamics of online violation and coercion. S2’s case, although framed as identity theft, also involved the non-consensual appropriation of her photos and identity, a violation that parallels the objectification and exploitation present in the Eid incident. All three participants were women of color, further aligning these cases with the intersectional dynamics of the reference scenario.

\subsubsection{Pairs' Perspective Restatement Quality via Similarity Analysis.}
\begin{figure*}
    \centering
    \includegraphics[width=.75\textwidth]{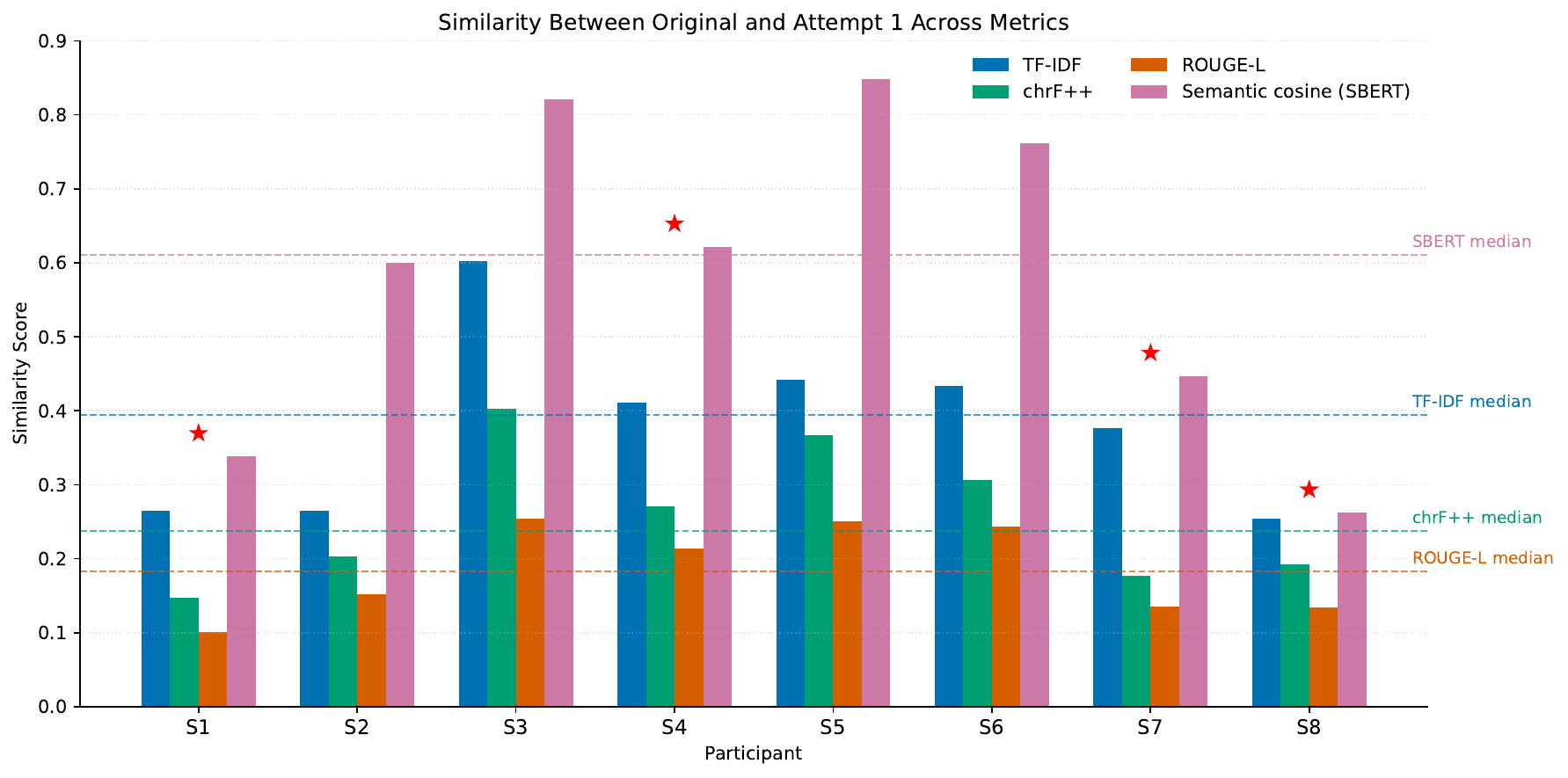}
    \caption{Similarity between participants’ restatements (Attempt~1) and the original user experience across four metrics (TF-IDF cosine, chrF++, ROUGE-L, and Semantic cosine via SBERT). Dashed lines mark the overall median for each metric. Red stars indicate participants who requested revisions to their Attempt~1.}
    \Description{Similarity between participants’ restatements (Attempt~1) and the original user experience across four metrics (TF-IDF cosine, chrF++, ROUGE-L, and Semantic cosine via SBERT). Dashed lines mark the overall median for each metric. Red stars indicate participants who requested revisions to their Attempt~1.}
    \label{fig:similarity}
\end{figure*}

To assess how faithfully participants re-articulated one another’s perspectives, we compared each participant’s initial restatement to the original user experience using TF-IDF cosine, chrF++, ROUGE-L, and SBERT. Figure~\ref{fig:similarity} visualizes these results with dashed lines marking each metric’s overall median (TF-IDF $\tilde{m}\!\approx\!0.394$, chrF++ $\tilde{m}\!\approx\!0.237$, ROUGE-L $\tilde{m}\!\approx\!0.183$, SBERT $\tilde{m}\!\approx\!0.611$) and red stars indicating cases where participants requested revisions.

We observed that three of the four participants who requested changes (S1, S7, S8) scored below the median on at least two of the three metrics, suggesting an alignment between lower measured similarity and participants’ own perceptions of inadequate perspective-taking. S4 represents a notable exception, requesting changes despite above-median similarity, indicating that textual metrics alone may not fully capture interpretive nuance or perceived fidelity. 

\noindent \textit{\textbf{Pair 1}}: S1’s scores are uniformly low (TF-IDF $=0.265$, chrF++ $=0.147$, ROUGE-L $=0.101$, SBERT $=0.338$), indicating limited fidelity at lexical, structural, and semantic levels—consistent with S1’s request for changes. By contrast, S2 shows similarly low lexical overlap (TF-IDF $=0.265$) but a substantially higher SBERT score ($0.600$), suggesting paraphrastic preservation of meaning despite rewording; chrF++ ($0.203$) and ROUGE-L ($0.152$) also improve relative to S1. Still, all four scores remain below their medians, aligning with the interpretation of only partial fidelity (S2 did not request changes).

\noindent \textit{\textbf{Pair 2}}: S3 aligns strongly with the original across all metrics (TF-IDF $=0.603$, chrF++ $=0.402$, ROUGE-L $=0.254$, SBERT $=0.821$), indicating high lexical, structural, and semantic faithfulness. S4 is also above median on every metric (TF-IDF $=0.411$, chrF++ $=0.271$, ROUGE-L $=0.214$, SBERT $=0.622$), yet still requested changes, underscoring that even semantically similar restatements may miss emphasis or nuance not captured by automated metrics.

\noindent \textit{\textbf{Pair 3}}: Both participants achieve above-median alignment across the board. S5 (TF-IDF $=0.441$, chrF++ $=0.367$, ROUGE-L $=0.250$, SBERT $=0.849$) and S6 (TF-IDF $=0.434$, chrF++ $=0.306$, ROUGE-L $=0.243$, SBERT $=0.762$) show strong fidelity; neither requested revisions, reinforcing the quantitative signal.

\noindent \textit{\textbf{Pair 4}}: Despite longer turns, S7’s scores are mixed (TF-IDF $=0.376$, chrF++ $=0.177$, ROUGE-L $=0.136$, SBERT $=0.447$), suggesting moderate semantic relatedness but weak phrasing/structural preservation. S8 shows uniformly low alignment (TF-IDF $=0.254$, chrF++ $=0.192$, ROUGE-L $=0.134$, SBERT $=0.262$). Both requested changes, illustrating that verbosity does not guarantee fidelity and that semantic similarity can remain low even with topical overlap.

\subsubsection{Contrasting Human–Human and Human–AI Perspective-Taking}
\begin{table}[ht]
\centering
\footnotesize
\caption{Overall increase in similarity across attempts (Human--AI). Top: change from Attempt~1 to the \textbf{best} attempt. Bottom: change from Attempt~1 to the \textbf{final} attempt. DoM = Difference of Means; MoD = Mean of Differences.}
\label{tab:trend_attempts}
\begin{tabular}{l l r r r r r}
\toprule
\textbf{Metric} & \textbf{Trend} & \textbf{Slope} & \textbf{DoM $\Delta$\%} & \textbf{MoD $\Delta$} & \textbf{MoD $\Delta$\%} & \textbf{Frac$\uparrow$} \\
\midrule
\multicolumn{7}{c}{\textit{Attempt 1 $\rightarrow$ Best Attempt}} \\
TF-IDF    &  $\uparrow$ & 0.0386  & 27.2\%  & 0.1254  & 55.5\%  & 0.9474 \\
chrF++    &  $\uparrow$ & 0.0182  & 30.0\%  & 0.0647  & 68.6\%  & 0.8947 \\
ROUGE-L   &  $\uparrow$ & 0.0085  & 18.3\%  & 0.0379  & 54.6\%  & 0.9474 \\
SBERT  &  $\uparrow$ & 0.0352   & 6.4\%   & 0.0715  & 14.9\%  & 0.6842 \\
\midrule
\multicolumn{7}{c}{\textit{Attempt 1 $\rightarrow$ Final Attempt}} \\
TF-IDF    &  $\uparrow$   & 0.0153  & 27.1\%  & 0.0921  & 43.2\%  & 0.7368 \\
chrF++    &  $\uparrow$   & 0.0027   & 7.8\%  & 0.0434  & 56.1\%  & 0.6842 \\
ROUGE-L   &  $\uparrow$   & 0.0033  & 17.9\%  & 0.0206  & 39.4\%  & 0.7895 \\
SBERT  &  $\downarrow$ & -0.0180 & -10.8\% & -0.0066 & 1.1\%   & 0.4737 \\
\bottomrule
\end{tabular}
\label{table:percent_change}
\end{table}

Overall, similarity scores in the human–AI condition showed clear improvement across attempts, particularly when comparing Attempt 1 to the best-performing attempt (Table \ref{table:percent_change}). All four metrics exhibited increases over time, with the strongest relative gains in chrF++ (+30.0\% DoM, +68.6\% MoD) and TF–IDF (+27.2\% DoM, +55.5\% MoD). Even semantic similarity improved by 6.4\% (DoM) and 14.9\% (MoD), with nearly 69\% of participants showing higher semantic alignment in their best attempt compared to their first. In contrast, when comparing Attempt 1 to the final attempt, gains were smaller and less consistent. TF–IDF and ROUGE-L continued to increase overall (+27.1\% and +17.9\%, respectively), but semantic similarity showed a slight overall decline (–10.8\% DoM), indicating that the best attempt often occurred before the final one. Taken together, these results suggest that participants generally refined their responses over multiple attempts, with most achieving their peak similarity prior to the final turn.

\begin{figure}
    \centering
    \includegraphics[width=\linewidth]{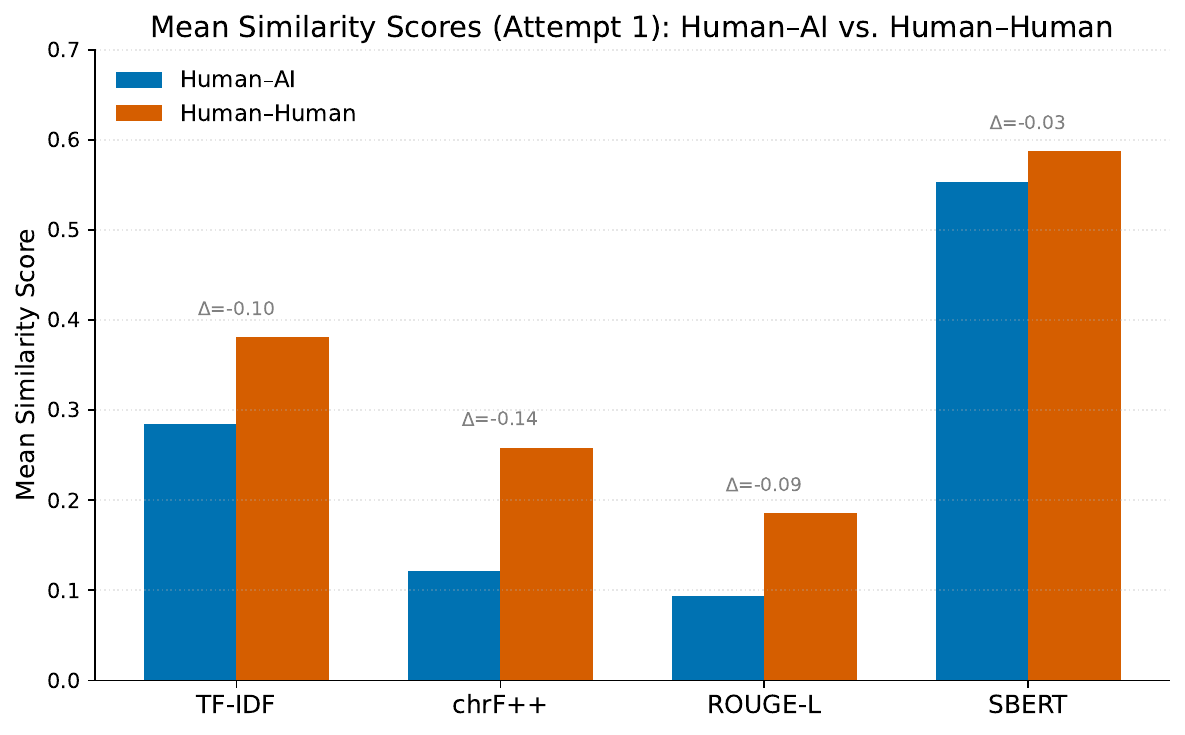}
    \caption{Mean similarity scores (Attempt 1) for Human–AI vs. Human–Human across four metrics. $\Delta$ above each pair shows $HA - HH$. }
    \Description{Mean similarity scores (Attempt 1) for Human–AI vs. Human–Human across four metrics. $\Delta$ above each pair shows $HA - HH$. Human-human exceeds human-ai across all metrics. }
    \label{fig:hhvhai}
\end{figure}

A comparison of mean similarity scores between the human–AI and human–human conditions for Attempt 1 (Figure \ref{fig:hhvhai}) reveals that the latter achieved consistently higher alignment with the original narratives across all metrics. Human–human conversations had higher mean scores for TF–IDF (0.3810 vs.\ 0.2845), chrF++ (0.2581 vs.\ 0.1210), ROUGE-L (0.1854 vs.\ 0.0936), and semantic similarity (0.5875 vs.\ 0.5529). These differences (ranging from –0.0346 for semantic similarity to –0.1371 for chrF++) indicate that participants working with another human started from a stronger baseline alignment with the original perspective than those interacting with the AI.

\subsubsection{Pairs' Conversation Quality via Turn-Taking Analysis}
\begin{table}[h!]
\centering
\footnotesize
\caption{Turn-based participation metrics (turns, median/mean words per turn, total words). Starred pairs* have a formal working relationship.}
\label{tab:turn_stats}
\begin{tabular}{llrrrrl}
\toprule
\textbf{Pair} & \textbf{Speaker demographics} & \textbf{Turns} & \textbf{Med.} & \textbf{Mean} & \textbf{Total}  \\
\midrule
P1 & S1 (white male asst. prof) & 58 & 6.0 & 20.8 & 1204 \\
 & S2 (POC female PhD)     & 53 & 6.0 & 8.5  & 449  \\
\midrule
P2*  & S3 (POC male PhD) & 46 & 13.5 & 20.0 & 919 \\
 & S4 (white male asst. prof)         & 47 & 9.0  & 14.8 & 696 \\
\midrule
P3 & S5 (POC female senior PhD) & 53 & 11.0 & 14.5 & 766 \\
 & S6 (POC female junior PhD) & 52 & 7.0 & 10.1 & 523 \\
\midrule
P4* & S7 (POC male PhD) & 54 & 14.5 & 27.2 & 1469 \\
 & S8 (white male asso. professor)     & 56 & 9.5  & 17.5 & 978  \\
\bottomrule
\end{tabular}
\label{table:hh_results}
\end{table}
Our goal was to contrast human–human perspective-taking with human–AI use of the Perspective Coach. A key contextual difference is turn-taking: in human–AI sessions the bot always responds and typically with long, informational turns; in human–human pairs, turn length and elaboration depend on roles, familiarity, and power.
Across Pairs 1–4 (human–human), floor access (turn counts) was broadly balanced, but turn length varied with relationship and role, see Table~\ref{table:hh_results} for details.

\noindent\textit{\textbf{Pair 1}}: Balanced turns but asymmetric elaboration: the higher-status participant produced much longer turns (mean 20.8 vs. 8.5 words; 1204 vs. 449 words total, see Table~\ref{table:hh_results}). This mirrors the risk in field settings: the developer side often sets the agenda and elaborates, while the community member contributes shorter responses. 

\noindent\textit{\textbf{Pair 2:}} Near-equal turn counts, but the junior produced longer turns (mean 20.0 vs. 14.8; 919 vs. 696 words). A short, explicit working relationship appeared to enable the junior’s elaboration, consistent with a strong working relationship.

\noindent\textit{\textbf{Pair 3:}} Balanced turns; the more senior participant produced longer turns (14.5 vs. 10.1 mean; 766 vs. 523 words), suggesting that seniority correlates with elaboration when identity is held constant.

\noindent\textit{\textbf{Pair 4:}} Balanced turns; the junior produced markedly longer turns (mean 27.2 vs. 17.5; 1469 vs. 978 words), again consistent with senior-as-facilitator / junior-as-producer in a collaborative relationship.
\section{Discussion}

 \textbf{(\textbf{\textit{RQ1}})} Our findings indicate that \perspective{} meaningfully supports developers in reflecting on their design decisions and considering alternative perspectives, especially those from marginalized users. Participants rated the tool highly for its ability to deepen reflection (Q1, $M = 4.68$) and broaden perspective-taking (Q2, $M = 4.74$), and qualitative feedback shows how this unfolded in practice. Participants described the chatbot as \qquote{“ethically educative”} and appreciated its ability to break down complex ethical issues into actionable steps, help them \qquote{“organize complex emotions,”} and deepen emotional articulation. These findings suggest that structured conversational prompts can act as scaffolds for critical reflection and empathy work within software design contexts, areas where traditional ethics tools often fail due to abstraction or lack of relevance~\cite{wong2023seeing,sirur2018we,selbst2019fairness}.

However, the results also reveal important limitations. Participants noted moments of generic repetition and over-structuring. These findings suggest that deeper reflection is best supported when the tool dynamically adapts to users’ needs and avoids rigid, prescriptive flows. Our participants’ calls for more specific, example-grounded guidance and less generic repetition point toward \textbf{multiplicity over singularity}: we tie this desire to an opportunity to create a tool that convenes a range of perspectives rather than a single “neutral” proxy. We want to reinforce the importance of designing \perspective{} to center marginalized users’ voices without reducing them to static representations. As such, we propose two \textbf{design commitments} for future versions of \perspective{}: \textbf{(1)} add \textbf{\textit{rotating, contextually diverse accounts}} instead of single-scenario accounts for each marginalized group, addressing participant feedback about generic, redundant responses while exemplifying the dimensionality of marginalized needs, \textbf{(2)} \textbf{\textit{prompt users to log concrete design changes}} and revisit them in future sessions, reflecting participant calls for practical, example-driven suggestions that link reflection to action.
    
\textbf{ (\textbf{\textit{RQ2}})} \perspective{} was broadly perceived as usable and relevant to participants’ day-to-day design work. Participants expressed a strong likelihood of future use (Q5, $M = 4.68$) and found the tool’s feedback relevant (Q3, $M = 4.58$), underscoring its potential to integrate ethical reasoning into existing workflows rather than feeling like an external add-on. Participants also described the chatbot as a “writing coach” that improved the clarity and precision of their language, broke habits of impersonal writing, and helped them articulate values-based reasoning—skills critical for documenting ethical justifications in software projects.

At the same time, participants highlighted areas where usability could improve. Some wanted clearer onboarding, a more conversational tone, and the ability to diverge from the scripted flow. Others requested more personalized feedback and greater flexibility to explore tangents. These insights point toward several actionable design strategies: (1) provide an in-flow preface that sets expectations for how to interact and when to push for precision versus free association; (2) include a \emph{tone selector} (e.g., conversational, clinical, reflective) to tailor the experience; and (3) reduce redundancy through variation strategies that track conversational history and introduce new angles (alternate perspectives from the same community) rather than repeating earlier points. Participants also suggested a softer visual interface, optional voice interaction, and faster response cadence to make the experience feel more like collaborative coaching than evaluation.

Together, these findings indicate that \perspective{} is already perceived as a usable and valuable tool for integrating ethical reflection into development practice. However, enhancing adaptivity, personalization, and user agency will be key to improving its relevance and sustaining engagement across diverse workflows and user preferences.

\textbf{(\textbf{\textit{RQ3}})} Our analysis of textual similarity and conversational dynamics reveals both the potential and the current limitations of \perspective{} relative to human–human dialogue. On the positive side, participants in the human–AI condition significantly improved their restatements across attempts (Table~\ref{table:percent_change}). All four metrics showed gains from Attempt~1 to a later, \emph{best} attempt, including TF--IDF (+27.2\%), chrF++ (+30.0\%), ROUGE-L (+18.3\%), and SBERT (+6.4\%), indicating that structured interaction helped participants capture both surface-level and semantic aspects of users’ experiences over time. However, semantic similarity declined slightly when comparing Attempt~1 to the \emph{final} attempt (–10.8\%), suggesting that peak semantic alignment often occurred earlier. We asked participants to complete five rounds of interaction, but once \perspective{} assessed that the perspective was well-taken, later exchanges shifted from restatement to reflections on the developer's own positionality and technical interventions. For example, after they took the perspective well, one participant told \perspective{}, “\qquote{It makes me question everything: how many of our ‘neutral’ design choices […] actively enable this?}” Another reflected, “\qquote{It makes me rethink whether, as a developer, I’m really doing something beneficial to society […] It redirects my thinking toward creating safer platforms for everyone, especially for women.}” 

The comparison with human–human sessions further contextualizes these results. Participants in human conversations started from a stronger baseline across all metrics, suggesting that the interpersonal context (i.e., social stakes, cognitive demands, task structure) may help humans capture meaning more accurately from the outset. Moreover, conversational dynamics differed markedly: while turn counts in human–human pairs were broadly balanced, turn length often varied with status, role, and prior relationship. Research shows that power asymmetries constrain low-power speakers’ narrative freedom, producing both self-silencing (under-informing) and extractive speech (over-informing)~\cite{Dotson_2011,kukla2014performative,mckinney2016extracted,kim2025low}. In Pair~1, for example, the higher-status participant produced substantially longer turns, mirroring patterns documented in mixed-gender and power-asymmetric settings~\cite{karpowitz2012gender,anderson1998meta,zimmerman2008sex}.  These findings suggest that while the AI ensures marginalized perspectives are consistently voiced, it cannot replicate the mutual adjustment, negotiation, and context-sharing that characterize human conversation. However, Messeri’s \emph{Land of the Unreal} warns that affective technologies keep marginalized representations close and actual marginalized people far from design processes~\cite{messeri2024land}, while Nakamura cautions that empathy work can gratify users' moral self-image without shifting power~\cite{nakamura2020feeling}. We therefore propose two more \textbf{design commitments} for the next version of \perspective{}, \textbf{(3) \textit{a guided practice-mode}}, where developers rehearse conversational parity and space-redistribution for live conversation, addressing turn-taking asymmetries observed in the human–human study, where higher-status participants dominated elaboration. \textbf{(4)} \textbf{\textit{invite original posters}} or other marginalized stewards to create editable, revocable scenarios, responding to our finding that participant requests for revisions did not always align with text similarity scores.
\section{Limitations \& Future Work}
Our work is exploratory and should be interpreted with several limitations in mind. First, \perspective{} represents an early design probe rather than a production-ready system. Its conversational scaffolds, feedback strategies, and interaction flows were intentionally kept simple to test feasibility and user responses, rather than to maximize performance. Future versions should incorporate more adaptive feedback, tone-control, and mechanisms to support listening-oriented interaction.

Second, both studies involved small, specific samples. The human--AI study included 18 front-end developers from WEIRD, English-speaking countries. This purposeful focus on Global North developer perspectives limits generalizability; future work should explore whether these effects hold across larger and more demographically diverse samples. Further, the human--human study, involved only four pairs of participants drawn from our own research community. While this small and familiar sample constrains the breadth of perspectives we can capture, it also offers a distinctive benefit: because we know participants’ professional backgrounds, relationships, and communication styles firsthand, we can interpret their interactions and power dynamics with greater contextual sensitivity. Given extensive evidence regarding the conversational dynamics between  marginalized and dominant interlocutors, our aim is not to reprove those claims but to examine, in a human–human version of our task, what human–AI designs may gain or forgo when mediated by a conversational tool. Our study is a starting point for that task; greater insights will be gained through full transcript analysis of both studies. 

Third, our experimental tasks focused on a single scenario: the harassment and objectification of Muslim women on Eid. While this scenario was chosen to foreground a salient concern, it does not capture the full range of ethical issues developers encounter. Responses and effectiveness may differ when applied to concerns related to privacy, accessibility, labor, or other domains. Further, due to the use of Prolific, we must consider the possibility of desirability bias. On Prolific, participants’ compensation is tied to task approval, which may have incentivized responses perceived as good rather than genuine. \perspective{}'s reliance on OpenAI’s GPT platform introduces non-determinism, meaning identical prompts can yield slightly different outputs across sessions, supporting personalized reflection, but complicates replication and consistency. Next, our textual similarity metrics provide useful but incomplete proxies for perspective-taking quality. Similarly, our conversational metrics (turn counts, verbosity) are indirect indicators of listening and engagement. Qualitative analysis of conversational content and interactional dynamics could complement these measures in future work. One human-human session was conducted online, which can subtly reshape conversational flow (longer between-turn gaps and fewer overlaps due to latency)~\cite{seuren2021whose, edwards2025impacts, tian2024corpus}; however, comparative studies of therapeutic sessions often find broadly similar engagement and relational outcomes between video and face-to-face settings~\cite{peasgood2023randomised, seuling2024therapeutic, fernandez2021live, greenwood2022telehealth}. Given our task emphasized re-articulation and mutual validation, and only one pair met online, we expect limited bias.

Finally, as an exploratory study, we examined tool use in a controlled, single-session context. We did not assess how perspective-taking practices or epistemic humility evolve over time, nor how \perspective{} might integrate into real-world software development workflows. Longitudinal deployments, more diverse datasets, and field-based evaluations are needed to understand the sustained impact and practical utility of such tools in professional settings.

\section{Conclusion}
By embedding plurality, accountability, and redistribution of conversational space at the center of developer practice, a future version of \perspective{} can move beyond performing understanding \emph{about} marginalized users toward facilitating understanding \emph{with} them. In doing so, it offers a starting point for reimagining how conversational AI might support more equitable design practices, not as a replacement for engagement, but as a tool that prepares developers to enter those conversations with greater humility, attentiveness, and care. This exploratory work points to the potential of LLM-based tools to help bridge persistent gaps between those who design technologies and those most affected by them.

\bibliographystyle{ACM-Reference-Format}
\bibliography{citations}

\end{document}